\makeatletter\def\input@path{
	{./graphics/}
}\makeatother%
\documentclass[twocolumn,preprintnumbers,pra,superscriptaddress]{revtex4}

\usepackage{amssymb,amsmath,graphicx,dcolumn,bm,float} 
\usepackage[utf8]{inputenc}
\usepackage[T1]{fontenc}
\usepackage{xcolor}
\usepackage[colorlinks,linkcolor=black,citecolor=blue]{hyperref}
\DeclareMathOperator{\sech}{sech}  
\begin{document}

\title{Synthetic magnetism for solitons in optomechanical array}
\author{P. Djorwé}
\email{djorwepp@gmail.com}
\affiliation{Department of Physics, Faculty of Science, 
University of Ngaoundere, P.O. Box 454, Ngaoundere, Cameroon}
\affiliation{Stellenbosch Institute for Advanced Study (STIAS), Wallenberg Research Centre at Stellenbosch University, Stellenbosch 7600, South Africa}

\author{H. Alphonse}
\email{ahouw220@yahoo.fr}
\affiliation{Department of Marine Engineering, Limb\'{e} Nautical Arts and Fisheries Institute, P.O Box 485, Limb\'{e}, Cameroon}

\author{S. Abbagari}
\email{abbagaris@yahoo.fr}
\affiliation{Department of Basic Science, National Advanced School of Mines and Petroleum Industries, The University of Maroua, P.O Box 08, ka\'{e}l\'{e}, Cameroon}

\author{S. Y. Doka}
\email{numami@gmail.com}
\affiliation{Department of Physics, Faculty of Science, 
University of Ngaoundere, P.O. Box 454, Ngaoundere, Cameroon}

\author{S. G. Nana Engo}
\email{serge.nana-engo@facsciences-uy1.cm}
\affiliation{Department of Physics, Faculty of Science, University of Yaounde I, P.O. Box 812, Yaounde, Cameroon}

\begin{abstract}
We propose a synthetic magnetism to generate and to control solitonic waves in $\rm{1D}-$optomechanical array. Each optomechanical cavity in the array couples to its neighbors through photon and phonon coupling. We create the synthetic magnetism by modulating the phonon hopping rate through a modulation frequency, and a modulation phase  between resonators at different sites. When the synthetic magnetism effect is not taken into account, the mechanical coupling play a crucial role of controlling and switching the waves from bright to dark solitons, and it even induces rogue wave-like a shape in the array. For enough mechanical coupling strength, the system enters into a strong coupling regime through splitting/crossing of solitonic waves, leading to multiple waves propagation in the array. Under the synthetic magnetism effect, the phase of the modulation enables a good control of the wave propagation, and it also switches soliton shape from bright to dark, and even induces rogue waves as well. Similarly to the mechanical coupling, the synthetic magnetism offers another flexible way to generate plethora of solitonic waves for specific purposes. This work opens new avenues for optomechanical platforms and sheets light on their potentiality of controlling and switching solitonic waves based on synthetic magnetism. 
\end{abstract}

\maketitle

\date{\today}


%
\section{Introduction} \label{Intro}
Optomechanical cavities (COM) have been revealed as being a nice platform for solitonic waves investigation owing to their intrinsic nonlinear interactions \cite{Leijssen_2017,Doolin_2014,Djorwe_2019,Kingni_2020,Djorw.2013}. Nonlinear interactions in simple COM have led to numerous interesting studies such as quantum entanglement \cite{Djorw.2014,Kotler_2021}, squeezing \cite{Djorwe_2013,Wollman_2015}, high-order sideband combs generation \cite{Cao_2016,Miri_2018,Djorwe_2020}, metamorphoses of basin boundaries \cite{Djorwe.2022}, and chaos \cite{Monifi_2016}. To harness more features related to optomechanical coupling, simple COMs have been extended to optomechanical array, and this has led to exotic physics including synchronization of COMs \cite{Djorwe.2020,Sheng_2020}, Parity-Time symmetry \cite{Ganainy_2018}, sensors based on exceptional points \cite{Hodaei_2017, Chen_2017, Djorwe.2019}, solitonic waves \cite{Alphonse_2022},  and bosonic synthetic magnetism for topological transport phenomena \cite{Fang_2012,Schmidt.2015,Fang_2017,Brendel.2017,Wang_2020}.  

Unlike electrons, photons and phonons do not feel a magnetic field,
because they are not charged. As such,  most of the physics connected with charged particles in magnetic fields is missing for bosons. To feed this gap, researchers started  to engineer artificial magnetic fields for bosons to make them akin to electrons in a magnetic field. Different schemes for synthetic magnetism have been proposed in the literature for specific purposes. These proposals are mostly related to array of COMs, where cells are connected through photon and/or phonon coupling. The engineering of the synthetic field lies on the modulation of this photon/phonon hopping rate. In \cite{Schmidt.2015}, an artificial magnetic field for photons  has been created in an optomechanical crystal. This magnetic field results from a modulation of photon coupling on the lattice, and was used to achieve photon transport in the presence of an artificial Lorentz force, edge states, and the photonic Aharonov–Bohm effect. Pseudomagnetic field for sound has been created in \cite{Brendel.2017} for  chiral transport of sound waves at nanoscale.  The resulted pseudomagnetic field was created by simple geometrical modifications of a snowflake phononic crystal. Along the same line, a magnetic gauge field for phonons has been demonstrated in a sliced optomechanical crystal nanobeam in \cite{Mathew_2020}. The gauge field were induced through multimode optomechanical interactions, and has been used to break time-reversal symmetry. Under a dynamical modulation of the system, the authors observed nonreciprocal phonon transport between resonators of different frequencies mediated by radiation pressure forces. In \cite{Wang_2020}, a novel mechanism to generate synthetic magnetic field for phonon lattice by Floquet engineering auxiliary qubits has been proposed to realize circulator transport and analog Aharonov–Bohm effects for acoustic waves. Owing to these interesting physics fostered by synthetic magnetism in optomechanics, here we propose to investigate solitonic waves in 1D optomechanical array which hosts both photon and phonon coupling between cells. In our proposal, the phonon coupling between cells is harmonically modulated in time, and this induces an artificial magnetic field for phonons. Such a mechanism has been realized in \cite{Fang_2012} to create an effective magnetic field for photons by controlling the phase of the dynamical modulation. By tuning the phase of the modulation, we observed (i) a perfect control and switching between bright and dark mechanical solitons along the array, (ii) a strong coupling regime inducing rogue waves, and (iii) multiple solitons crossings/splittings. These interesting features reveal the crucial effect of the synthetic magnetism on the solitonic waves, and can be extended to similar systems. This work opens new avenues for solitons generation, and provides a smooth way to switch from bright to dark solitons depending on the purposes. 

The rest of this work is organized as follows. In \autoref{sec:MEQ}, we describe the system and the related equations. The mechanical coupling and the synthetic magnetism effect on the solitonic waves are presented in \autoref{sec:Sol.Dyn} and we conclude the work in \autoref{sec:Concl}.

\section{Model and dynamical equations} \label{sec:MEQ}

\begin{figure}[tbh]
  \centering
  \includegraphics[width=.9\columnwidth]{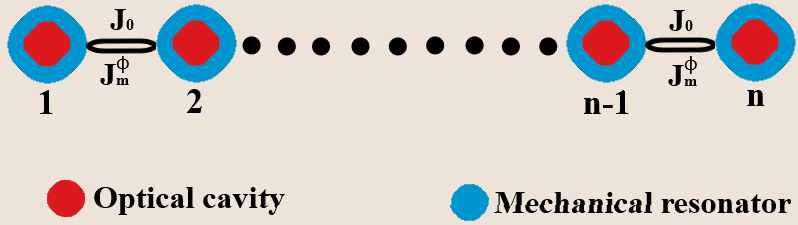}
  \caption{Dynamically modulated $\rm{1D}-$optomechanical array for an
effective magnetic field for phonons. Each cell supports a mechanical resonator (blue color) and an optical cavity (red color). The neighbors' cell are optically (mechanically) coupled through the hopping coupling rate $J_0$ ($J_m$). The mechanically hopping rate is dynamically modulated as $J_m (\phi)=J_m \cos{(\Omega t +\phi)}$.}
  \label{fig:Setup}
  \end{figure}

Our proposal consists of $N$ cells of COM which are both optically and mechanically coupled to each other, as sketched in \autoref{fig:Setup}.
The mechanical resonator and the optical cavity of each cell are depicted by the blue and red color respectively, and the photon (phonon) hopping coupling rate between neighboring cells is captured by $J_0$ ($J_m$). The dynamics of such an array of COM is captured by the Hamiltonian (with $\hslash=1$),
\begin{equation}\label{eq:Hamiltonian}
H=H_0+H_{J_0}+H_{J_m}+H_{int},
\end{equation}
with
\begin{align}
H_0&=\sum_n\left(\omega_{m,n} b^{\dagger}_nb_n +\omega_{c,n} a^{\dagger}_na_n\right), \label{Eq:eq1}\\
H_{J_0}&=- \frac{J_0}{2}\sum_n \left(a^{\dagger}_na_{n+1}+a_n a^{\dagger}_{n+1}\right), \label{Eq:eq2}\\
H_{J_m}&=- \frac{J_m (\phi)}{2}\sum_n\left(b^{\dagger}_nb_{n+1}+b_n b^{\dagger}_{n+1}\right), \label{Eq:eq3}\\
H_{int}&=-\sum_n g_n a^{\dagger}_n a_n\left(b^{\dagger}_n+b_n\right), \label{Eq:eq4}
\end{align}
where $H_0$, $H_{J_0} (H_{J_m})$, and $H_{int}$ are the free optomechanical, the photon (phonon) hopping and optomechanical interaction Hamiltonian, respectively. The optical and mechanical mode are described by their $n^{th}$ related bosonic field operators,  $a_n$ ($a^{\dagger}_n$) and $b_n$ ($b^{\dagger}_n$), respectively. The mechanical displacement is connected to its operators through $x_n=x^n_{ZPF}(b_n+b^{\dagger}_n)$ where $x^n_{ZPF}$ is the zero-point fluctuation amplitude of the $n^{th}$ mechanical resonator. The other parameters are the mechanical (cavity) frequency $\omega_{m,n}$ ($\omega_{c,n}$), and the optomechanical coupling $g_n$. The term $J_m (\phi)$ in \autoref{Eq:eq3} is the modulated phonon hopping term, which is introduced to reveal the synthetic magnetism features. This term is explicitly given by $J_m (\phi)=J_m \cos{(\Omega t +\phi)}$, where $\Omega$ is the modulation frequency, and $\phi$ is the phase of the modulation between resonators
at different sites. From now on, we assume that our cells are degenerated, and therefore use a simplified notation by dropping the index $n$ such as $\omega_{m,n}\equiv\omega_m$, $\omega_{c,n}\equiv\omega_c$, and $g_n\equiv g$. From the Hamiltonian given in \autoref{eq:Hamiltonian}, one deduces the following Quantum Langevin Equations (QLEs),
\begin{equation}\label{Eq:eq5}
\begin{split}
\dot{a}_n&=i\left[-\omega_c+g(b^{\dagger}_n+b_n)\right]a_n+i\frac{J_0}{2}(a_{n+1}+a_{n-1}),\\
\dot{b}_n&=-i\omega_mb_n+ig|a_n|^2+i\frac{J_m (\phi)}{2}(b_{n+1}+b_{n-1}). 
\end{split}
\end{equation}
To capture the soliton dynamics from \autoref{Eq:eq5}, we assume that photon and phonon numbers are large enough to consider the classical limit where noise terms can be neglected. Moreover, the operators can be substituted by their $c$-mean values $\langle a_n\rangle=\alpha_n$ and $\langle b_n\rangle=\beta_n$. In the rotating frame moving at $\omega_c+J_0$, the above QLEs yields,
\begin{equation}\label{Eq:eq6}
\begin{split}
\dot{\alpha}_n&=i[g(\beta^{\dagger}_n+\beta_n)]\alpha_n+i\frac{J_0}{2}(\alpha_{n+1}+\alpha_{n-1}-2\alpha_n),\\
\dot{\beta}_n&=-i\omega_m\beta_n+ig|\alpha_n|^2+i\frac{J_m (\phi)}{2}(\beta_{n+1}+\beta_{n-1}-2\beta_n), 
\end{split}
\end{equation}
which is the set of equations that will be considered later on in our numerical simulations. 

\section{Soliton dynamics under the mechanical coupling and the synthetic magnetism} \label{sec:Sol.Dyn}

To carry out numerical analysis of our proposal, we consider the dynamical equations given in \autoref{Eq:eq6}. We seek to harness the synthetic magnetic field effect on the solitonic wave propagation under the initial solution, 
 \begin{equation}\label{Eq:eq7}
  \alpha_n(0)=\tfrac{\sqrt{\frac{\omega_m J_0}{2}}}{gx_0}\sech\left(\frac{n-x(0)}{x_0}\right)\exp(i\rm{k}n),
 \end{equation}
 where $x_0$ and $x(0)$ are the soliton width and the initial position of the envelope respectively, and $\rm{k}$ is the wave vector. Throughout the numerical simulations, the initial conditions $\alpha_n=\alpha_n(0)$ and $\beta_n(0)=0$ will be used for the optical field and the mechanical resonator, respectively. Since we are interested on the mechanical coupling and synthetic magnetism effects, the wave vector $\rm{k}$ will be set to zero ($\rm{k=0}$).
 
 \subsection{Soliton dynamics under the mechanical coupling effects} \label{sec:Sol.DynI}

\begin{figure}[tbh]
  \centering
  \includegraphics[width=.95\columnwidth]{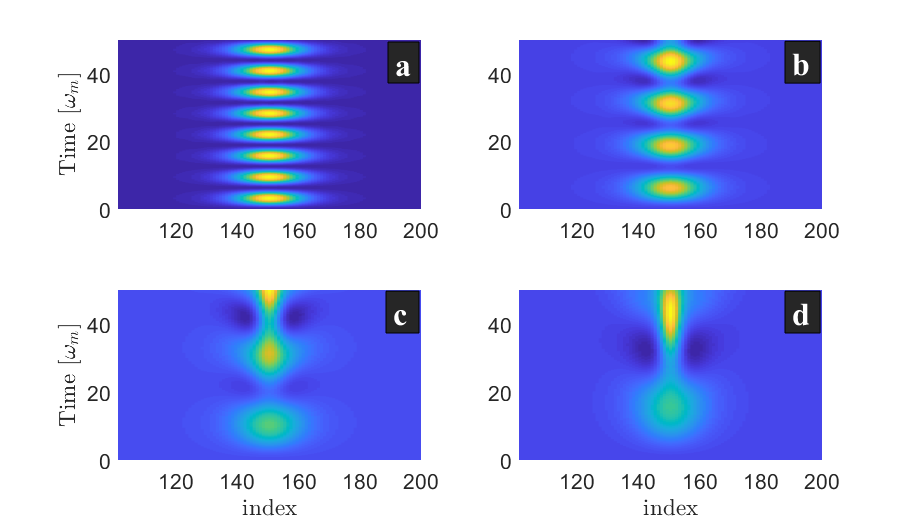}
  \caption{Dynamics of bright solitons in index and time $\rm{2D}$ representation. The mechanical coupling from (a) to (d) are $\tfrac{J_m}{\omega_m}=0; 0.5; 0.7$ and $0.8$, respectively. The rest of the parameters are $N=100$, $J_0=2\omega_m$, $g=10^{-4}\omega_m$, $x_0=10$, $n=100$, $x(0)=\frac{n}{2}$, $\Omega=0$, $\phi=0$ and $k=0$.}
  \label{fig:Mecha1}
  \end{figure}

\begin{figure}[tbh]
  \centering
  \includegraphics[width=.95\columnwidth]{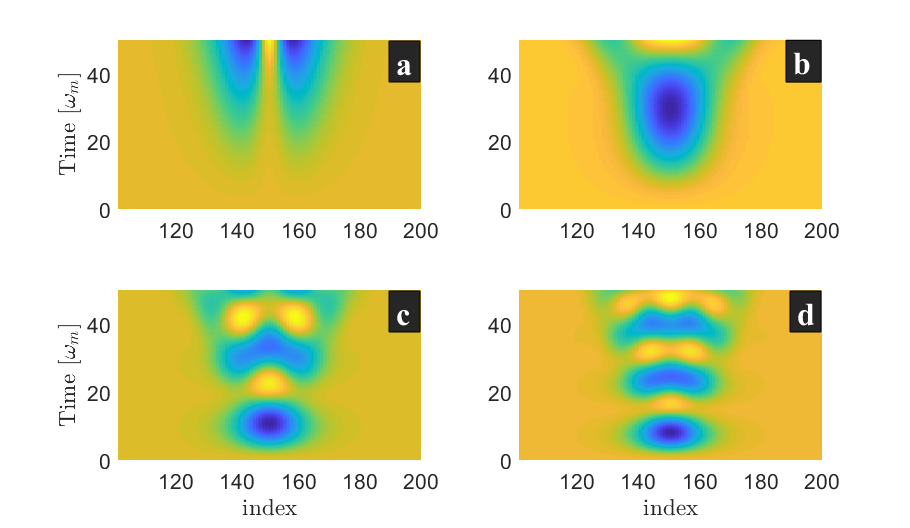}
  \caption{Dynamics of dark solitons in index and time $\rm{2D}$ representation. The mechanical coupling from (a) to (d) are $\tfrac{J_m}{\omega_m}=1.01; 1.1; 1.3$ and $1.4$, respectively. The other parameters are the same as in \autoref{fig:Mecha1}.}
  \label{fig:Mecha2}
  \end{figure}

We are interesting here in analyzing the effects of the mechanical coupling on solitonic waves. Therefore, the synthetic magnetism effect is not accounted ($\Omega=0; \phi=0$).  \autoref{fig:Mecha1} shows the $\rm{2D}$ projection profile of mechanical soliton emerging in our system. These figures correspond to different values of the mechanical coupling $J_m$ when the effect of the synthetic magnetism is not taken into account. The values of $J_m$ for this set of figures are $\tfrac{J_m}{\omega_m}=0; 0.5; 0.7$ and $0.8$ for \autoref{fig:Mecha1} ((a); (b); (c) and (d)), respectively. Different shapes of the mechanical soliton can be seen, and all are propagating at the center of the array.
 It can be observed that there are more and stable solitonic harmonics propagating in \autoref{fig:Mecha1}(a) than in \autoref{fig:Mecha1}(b). In \autoref{fig:Mecha1}(c) and \autoref{fig:Mecha1}(d), the soliton peaks are less present on the array, and their high is also decreasing compared to those in \autoref{fig:Mecha1}(a) and \autoref{fig:Mecha1}(b). More importantly, one observes that some hollows/deeps are appearing on \autoref{fig:Mecha1}(c) and \autoref{fig:Mecha1}(d). These hollows reveal the fact that there are new solitons which are propagating with reversed peaks in the array, and such a feature is reminiscent of dark solitons. This transition from bright to dark solitonic waves can be further pointed out by keeping increase the mechanical coupling, as depicted in \autoref{fig:Mecha2}. From \autoref{fig:Mecha2}(a) to \autoref{fig:Mecha2}(d), the phonon hopping rate is ramping from $\tfrac{J_m}{\omega_m}=1.01; 1.1; 1.3$ to $1.4$.
 
\begin{figure}[tbh]
  \centering
  \includegraphics[width=.95\columnwidth]{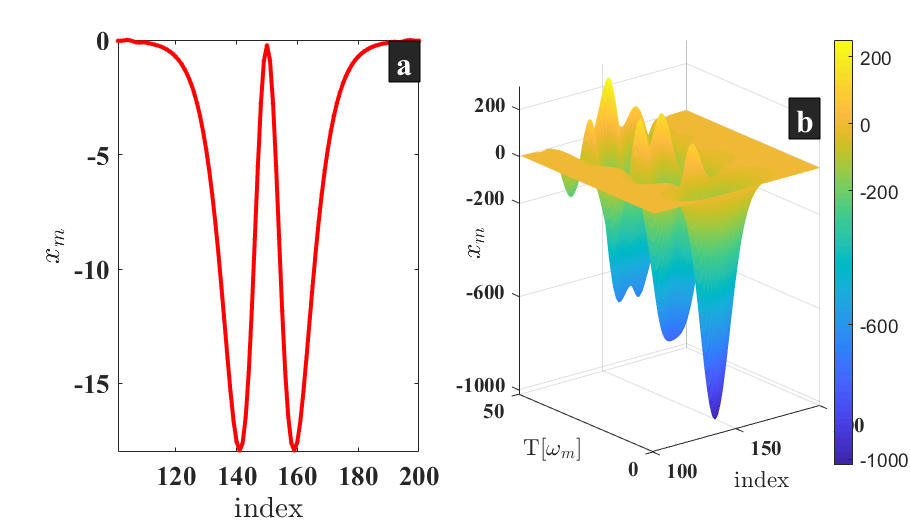}
  \caption{(a) is the representation of \autoref{fig:Mecha2}(a) at $t=50\omega_m$ to highlight the phase transition between bright and dark soliton. (b) is a $\rm{3D}$ illustration of \autoref{fig:Mecha2}(d) to depict rogue waves feature in strong coupling regime ($J_m=1.4\omega_m$).}
  \label{fig:Mecha3D}
  \end{figure}

By comparing \autoref{fig:Mecha1} and \autoref{fig:Mecha2}, one observe that the background of these set of figures has switched from dark to bright, meaning that the peaks on these two configurations are globally reversed one to another, and this reveals that the transitional phase from bright to dark solitonic waves has been realized. In \autoref{fig:Mecha2}(a), it can be seen that the increase of $J_m$ has led to the generation of two separated dark solitons. As this coupling keep increase, it induces the crossing/splitting of these two dark solitons as shown in \autoref{fig:Mecha2}(b), which means that the system has reached a strong coupling regime. For a strong enough strength of $J_m$, the array exhibits  complex features as depicted in \autoref{fig:Mecha2}(c) and \autoref{fig:Mecha2}(d). These complex features seem to exhibit a combination of both bright and dark solitons, which is a signature of rogue waves. To highlight the transitional phase between bright/dark solitonic wave and the rogue wave features, we have exemplified some aspects of \autoref{fig:Mecha2}(a) and \autoref{fig:Mecha2}(b) in \autoref{fig:Mecha3D}. 

\autoref{fig:Mecha3D}(a) is a representation of \autoref{fig:Mecha2}(a) at the longest time considered here ($t=t_{max}=50\omega_m$). This figure shows how the maximum peak of the soliton is around zero, which agrees with the observation made in \autoref{fig:Mecha1} pointing out that an increase of the mechanical coupling $J_m$ induces a decrease of the soliton's dark's peak. Moreover, the two hollows/deeps which started from \autoref{fig:Mecha1}(c) (and amplified in \autoref{fig:Mecha2}(a)) are clearly illustrated in \autoref{fig:Mecha3D}(a) which shows how the solitonic wave is at the transitional threshold between bright to dark solitons. Indeed, increasing $J_m$ will totally lead to negative soliton's peak. We conclude that this phase transition happens at the vicinity of $J_m=\omega_m$. \autoref{fig:Mecha3D}(b) is a $\rm{3D}$-representation of \autoref{fig:Mecha2}(d) (for $J_m=1.4\omega_m$), which is well above the transitional threshold. As predicted, this figure depicts a mixing of bright and dark solitons which is a rogue waves signature. Owing to these features, we conclude that the mechanical coupling strength can be used to control solitonic waves propagation and to switch from bright to dark solitons in an optomechanical array.

\begin{figure}[tbhp]
  \centering
  \includegraphics[width=.95\columnwidth]{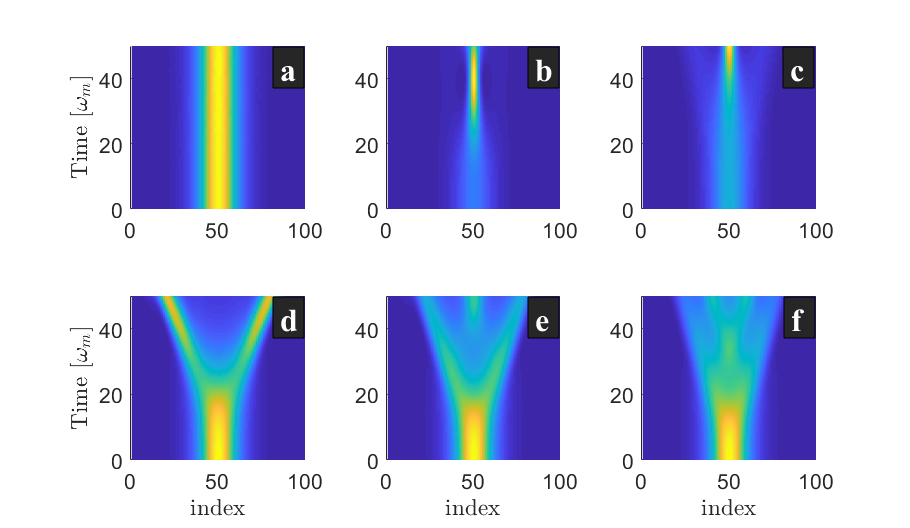}
  \caption{Dynamics of the optical solitonic waves, showing solitonic waves splittings/crossings due to a strong coupling. From (a) to (f), the mechanical coupling is respectively, $\tfrac{J_m}{\omega_m}=0; 0.8; 1.01$, $1.1$, $1.2$ and $1.3$. The other parameters are the same as in \autoref{fig:Mecha1}. One can see two (c), three (d) and four (e) solitonic waves crossing/splitting in the array.}
  \label{fig:Opt}
  \end{figure}
  
To point out the optomechanical nature of our study, we carry out the effect of the mechanical coupling on the optical solitons as well. This analysis is shown in \autoref{fig:Opt} where the optical coupling strength is maintained at $J_0=2\omega_m$. As it can be seen, there is a stable optical solitonic wave that is propagating at the center of the array in \autoref{fig:Opt}(a), which corresponds to the case without mechanical coupling $J_m=0$. The effect of the phonon hopping can be seen from \autoref{fig:Opt}(b), where the width of the wave is shrinking, and it becomes blurred in \autoref{fig:Opt}(c) as the phonon hopping rate increases from $\tfrac{J_m}{\omega_m} =0.8$ to $1.01$. This blurred shape of the optical solitons is the sign that their peak is decreasing,  meaning that the wave is losing its energy that is spread to cells located far from the array's center. To confirm such statement, we have further increases phonon hopping strength ($\tfrac{J_m}{\omega_m} =1.1$), and we observed an optical soliton splitting in \autoref{fig:Opt}(c). This splitting reveals that our system has reached a strong coupling regime, where solitons can cross or split, leading to multiple solitonic waves propagating in the array. Indeed, \autoref{fig:Opt}(c) to \autoref{fig:Opt}e are plotted respectively for $\tfrac{J_m}{\omega_m} =1.1; 1.2; 1.3$, which clearly are above the strong coupling regime, and we can see two (see \autoref{fig:Opt}(c)), three (see \autoref{fig:Opt}(d)) and four (see \autoref{fig:Opt}e) solitonic waves propagating in the array. From this analysis, we conclude that the strong coupling induced by the mechanical coupling is reflected both in our optical and mechanical resonators through the splitting/crossing of the solitonic waves. This interplay between optical and mechanical waves in our array is mediated by the optomechanical coupling $g$, which is the key parameter that connects optical and mechanical resonators in each cell of the array.         

\subsection{Soliton dynamics under the synthetic magnetism effect} \label{sec:Sol.DynII}

To point out the effect of the synthetic magnetism, we can either set on the modulation
frequency $\Omega$ or the phase of the modulation between resonators at different sites $\phi$. For a proof of concept purpose and without lost of generality, we will not set on these two parameters simultaneously in this work, since they have similar effects. However, we will mostly focus on the phase-induced effects. Moreover, we would like to point out that (i) for $(\Omega t +\phi)\equiv \frac{n\pi}{2}$ with $n\in \mathbb{Z^{\ast}}$, our system only hosts the optical coupling between cells and the related results have been shown in \cite{Alphonse_2022}; (ii) for $(\Omega t +\phi)\equiv n\pi$ with $n\in \mathbb{Z}$, our proposal does not host synthetic magnetism and the related results reduce to those shown in \autoref{fig:Mecha1} and \autoref{fig:Mecha2}. 

\begin{figure}[tbh]
  \centering
  \includegraphics[width=\columnwidth]{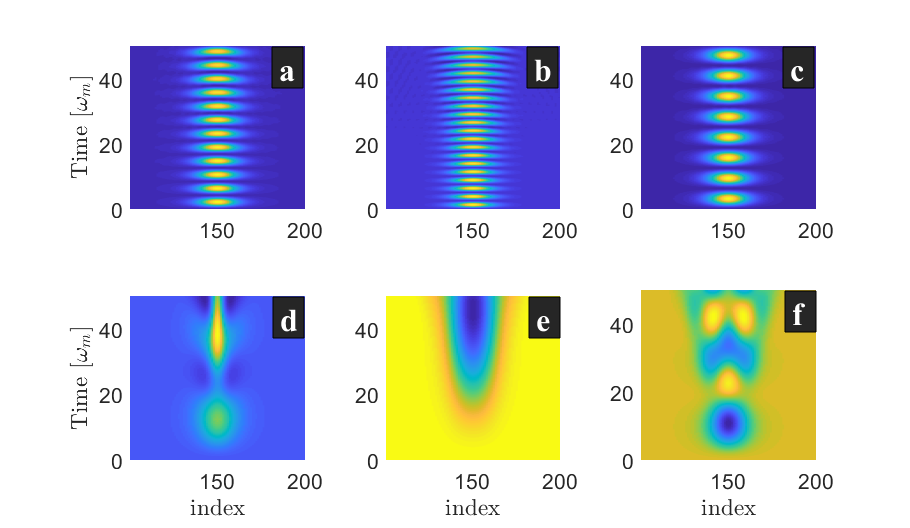}
  \caption{Synthetic magnetism effect on the soliton dynamics. (a) corresponds to $\Omega=\tfrac{2\pi\nu}{3}$ and $\phi=0$ while in (b) to (f), $\Omega=0$ and the phase is respectively fixed to $\phi=\pi; \frac{\pi}{2}; \frac{\pi}{3}; \frac{\pi}{4}$ and $\frac{\pi}{6}$. The mechanical coupling is fixed to $\tfrac{J_m}{\omega_m}=0.5$ for (a)-(c) and to $\tfrac{J_m}{\omega_m}=1.5$ for (d)-(e). The other parameters are the same as in \autoref{fig:Mecha1}.}
  \label{fig:Mecha3}
  \end{figure}

The effect of the synthetic magnetic field is depicted in \autoref{fig:Mecha3}, where the mechanical coupling is fixed to $\tfrac{J_m}{\omega_m}=0.5$ for \autoref{fig:Mecha3}(a-c) and to $\tfrac{J_m}{\omega_m}=1.5$ for \autoref{fig:Mecha3}(d-e). To point out the effect of the modulation frequency, we have tuned it to $\Omega=\tfrac{2\pi\nu}{n}$ where the  frequency is $\nu=50\rm{Hz}$, with $n \in \mathbb{Z^{\ast}}$. Starting from \autoref{fig:Mecha1}(b) and by setting $n=3$ and $\phi=0$, we obtained \autoref{fig:Mecha3}(a). It can be seen in \autoref{fig:Mecha3}(a) that the peaks have spread along the center of the array, meaning that $\Omega$ enhances propagation of solitonic waves. A similar effect is observed in \autoref{fig:Mecha3}(b), where we have set $\Omega=0$ and $\phi=\pi$. As we previously predicted, \autoref{fig:Mecha3}(c) depicts the solitonic wave propagation when the effect of the mechanical coupling is annihilated for $\phi=\frac{\pi}{2}$, leading to a well known result (see \autoref{fig:Mecha1}(a)). To further figure out the effect of the phase ($\phi$) on our system, let us still start with a stable configuration corresponding to $\Omega=0$, $\phi=0$ and $J_m=1.5 \omega_m$ (not shown). From such a stable solitonic wave, we tune the phase to $\phi=\frac{\pi}{3}; \frac{\pi}{4}$ and $\frac{\pi}{6}$. The corresponding outputs are respectively depicted on \autoref{fig:Mecha3}(d)-(e), and it is clearly seen that, like $J_m$, the phase induces a transition from bright to dark solitons. Furthermore, the rogue wave-like feature can be also recovered through phase adjustment. We can thus conclude that similar to the phonon hopping effect, the phase $\phi$ is equally an important parameter to control solitonic waves and to smoothly switch from bright to dark soliton features.       

\section{Conclusion} \label{sec:Concl}

This work investigated the effect of mechanical coupling and a synthetic magnetic field on solitonic waves propagating in $\rm{1D-}$optomechanical array. Within the array, each optomechanical cavity is coupled to its neighbors through photon and phonon couplings. The synthetic magnetism is created by modulating the phonon hopping rate. We have shown that the tuning of the mechanical coupling leads to several solitonic waves features ranging from bright to dark solitons, and even to the emergence of rogue waves in the array. Moreover, this phonon hopping strength induces a strong coupling regime revealed through splitting/crossing of solitonic waves leading to multiple waves propagating in the array. For a fixed strength of the mechanical coupling, we have also shown that the frequency and the phase of the synthetic magnetism are crucial parameters to control the solitonic wave as well. By focusing on the phase, we were able to not only control the wave propagation, but this phase has been proven to switch soliton shape from bright to dark, and even to induce rogue waves in the array. This work opens new avenues for optomechanical platforms and provides ingredients based on phononic coupling and synthetic magnetism to control and smoothly switch solitonic waves from one shape to another.

\bibliography{Sensor}

\end{document}